\documentclass[twocolumn,amsmath,amssymb,prb]{revtex4}

\usepackage{graphicx}
\usepackage{dcolumn}
\usepackage{bm}

\begin{document}

\title{Thermally driven spin injection from a ferromagnet into a non-magnetic metal}

\author{A. Slachter}
 \email{A.Slachter@rug.nl}
 \author{F. L. Bakker, J. -P. Adam}
\author{B. J. van Wees}%

\affiliation{Physics of Nanodevices, Zernike Institute for Advanced Materials, University
of Groningen, The Netherlands
}

\date{\today}

\maketitle

\textbf{Creating, manipulating and detecting spin polarized carriers are the key elements of spin based electronics\cite{zuticrevmodphys,chappertnatmat}. Most practical devices\cite{BaibichPRL,STO,CIMS} use a perpendicular geometry in which the spin currents, describing the transport of spin angular momentum, are accompanied by charge currents. In recent years, new sources of pure spin currents, i.e., without charge currents, have been demonstrated\cite{Jedema,Valenzuela,Costache,Saitoh} and applied\cite{Otani1,Otani2,Crowell,Dash}. In this paper, we demonstrate a conceptually new source of pure spin current driven by the flow of heat across a ferromagnetic/non-magnetic metal (FM/NM) interface. This spin current is generated because the Seebeck coefficient, which describes the generation of a voltage as a result of a temperature gradient, is spin dependent in a ferromagnet\cite{HatamiBauer, Tulapurkar}. For a detailed study of this new source of spins, it is measured in a non-local lateral geometry. We developed a 3D model that describes the heat, charge and spin transport in this geometry which allows us to quantify this process\cite{FLBakker}. We obtain a spin Seebeck coefficient for Permalloy of -3.8 $\mu$V/K demonstrating that thermally driven spin injection is a feasible alternative for electrical spin injection in, for example, spin transfer torque experiments\cite{HatamiPRL}.}

\begin{figure}[b]
\includegraphics[width=8.8cm,keepaspectratio=true]{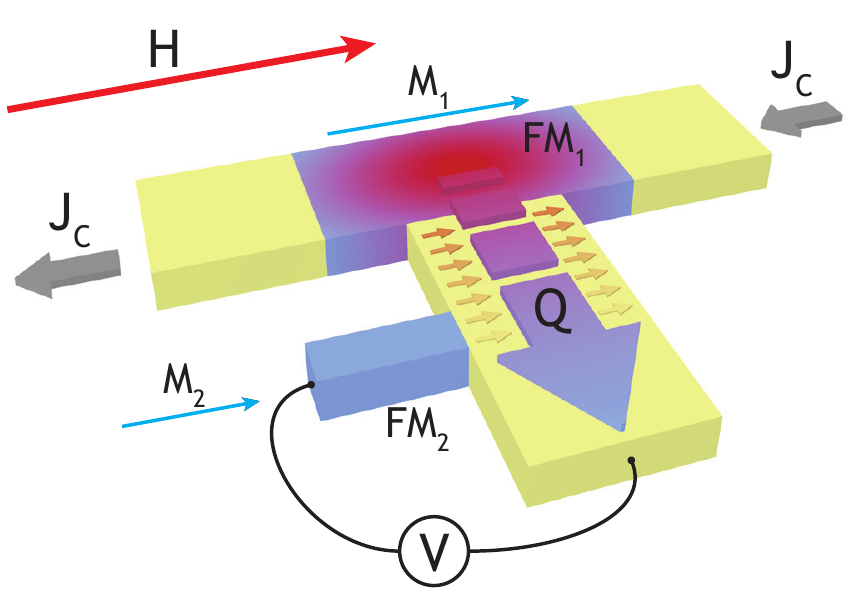}
\caption{\label{fig:1}\textbf{Conceptual diagram}. A charge current J$_{C}$ is sent through ferromagnet 1 (FM$_{1}$) causing Joule heating due to the large resistivity of FM$_{1}$. The NM contacts (yellow) are highly thermally conductive, thereby providing heat sinks. The heat current Q through the center FM$_{1}$/NM interface injects a spin current into the NM depending on the magnetization direction M$_{1}$. The generated spins diffuse towards the FM$_{2}$/NM interface where they generate a potential $\Delta \mu=P\mu_{s}$ depending on the magnetization direction M$_{2}$. As a consequence of Joule heating, the signal expected to arise from thermal spin injection scales with $\nabla T \propto I^{2}$. This potential is measured using the indicated voltage scheme by selectively switching the magnetization directions M$_{1}$ and M$_{2}$ by a magnetic field H.}
\end{figure}

The interplay of spin dependent conductivity and thermoelectricity was already known for half a century where it was used to describe the conventional Seebeck effect of ferromagnetic metals\cite{TEP72}. The discovery of the GMR effect\cite{BaibichPRL} sparked the interest of the community in spin dependent conductivity and novel spin electronics which is going on until today\cite{CIMS, STO, Costache, magndyn}. Due to experimental difficulties in controlling heat flows it was only until very recent that thermoelectric spintronics was investigated\cite{Gravier1, Gravier2,SpinSeebeckQD} leading to the new field of spin caloritronics\cite{HatamiBauer}. A relevant example is given by Uchida \textit{et al.\ }\cite{Saitoh} who interpreted their results in terms of the generation of a \textit{bulk} spin accumulation due to an applied temperature gradient in a ferromagnet film. In contrast, the effect we describe in this paper arises from a heat current flowing through a ferromagnetic/non magnetic metal junction (FM/NM) which creates a spin accumulation at the \textit{interface}.

The concept of how we generate a heat current over a FM/NM junction and subsequently measure the spin accumulation is shown in figure \ref{fig:1}. The scheme is essentially a lateral non local spin valve structure\cite{Jedema} with the electrical injection replaced by thermal spin injection. We use this non local scheme to separate spin injection from possible spurious effects\cite{Jedema,Dash,Crowell} and because the observed thermally generated non-spin related voltage, which we refer to as the baseline resistance, allows to extract the temperature distribution in the device by comparing this to modeling\cite{FLBakker}.

\begin{figure}[b]
\includegraphics[width=8.8cm,keepaspectratio=true]{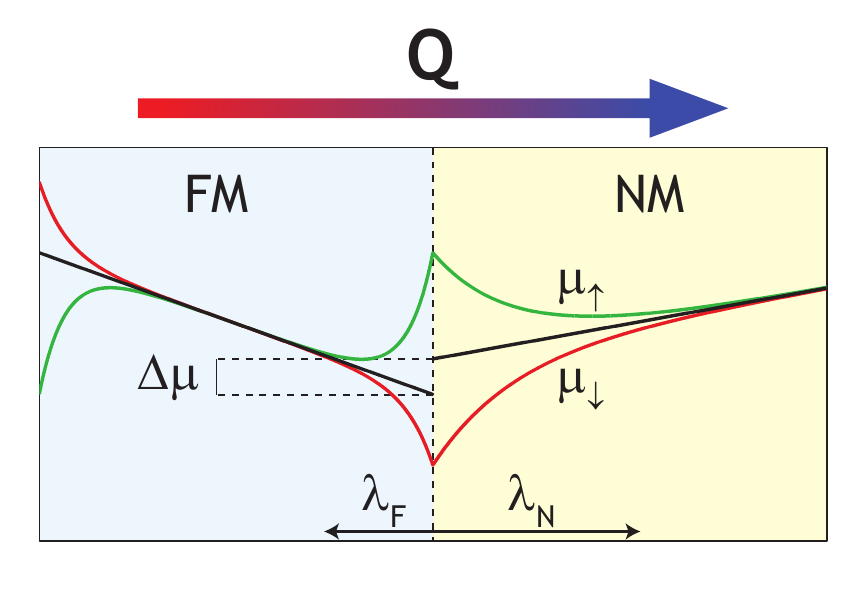}
\caption{\label{fig:2}\textbf{Thermal spin injection by the spin Seebeck coefficient across a FM/NM interface}. Schematic figure showing the resulting spin dependent chemical potentials $\mu_{\uparrow,\downarrow}$ across a FM/NM interface when a heat current $Q=-k\nabla T$ crosses it. Heat current is taken to be continuous across the interface leading to a discontinuity in $\nabla T$. No currents are allowed to leave the FM, nevertheless, a spin current proportional to the spin Seebeck coefficient flows through the bulk FM which needs to become unpolarized in the bulk NM. This creates a spin imbalance $\mu_{\uparrow}-\mu_{\downarrow}$ at the boundary which relaxes in the FM and NM on the length scale of their respective spin relaxation lengths $\lambda_{i}$. A thermoelectric interface potential $\Delta \mu=P\mu_{s}$ also builds up. On the left side no spin current is allowed to leave leading to a spin accumulation of opposite sign.}
\end{figure}

We first formulate an appropriate diffusive transport theory for thermally driven spin injection. The Seebeck coefficient describes that an applied temperature gradient across a conductor generates an electric field\cite{Mott}. In a ferromagnet, the transport processes for the majority and minority spin are different leading to a spin dependent conductivity $\sigma_{\uparrow,\downarrow}$ and Seebeck coefficient $S_{\uparrow,\downarrow}$\cite{TEP72,Tulapurkar}. The first is used to describe magnetoelectronics\cite{magnel} in FM/NM systems where the latter one is usually disregarded. In order to consider what happens when heat is sent through the system, we write the spin dependent currents:

\begin{equation}\label{eq:currents}
J_{\uparrow,\downarrow}=-\sigma_{\uparrow,\downarrow}(\frac{1}{e}\nabla \mu_{\uparrow,\downarrow}-S_{\uparrow,\downarrow}\nabla T)
\end{equation}

\noindent here $\mu_{\uparrow,\downarrow}$ is the spin dependent chemical potential. When a heat current Q is sent through a ferromagnet in the absence of a charge current, a spin current $J_{s}=J_{\uparrow}-J_{\downarrow}=\sigma_{F}(1-P^{2})S_{s}\nabla T/2$ flows driven by the spin Seebeck coefficient $S_{s}=S_{\uparrow}-S_{\downarrow}$. Here $P$ is the current polarization of the ferromagnet and $\sigma_{F}$ is the conductivity of the ferromagnet. 

\begin{figure}[b]
\includegraphics[width=8.8cm,keepaspectratio=true]{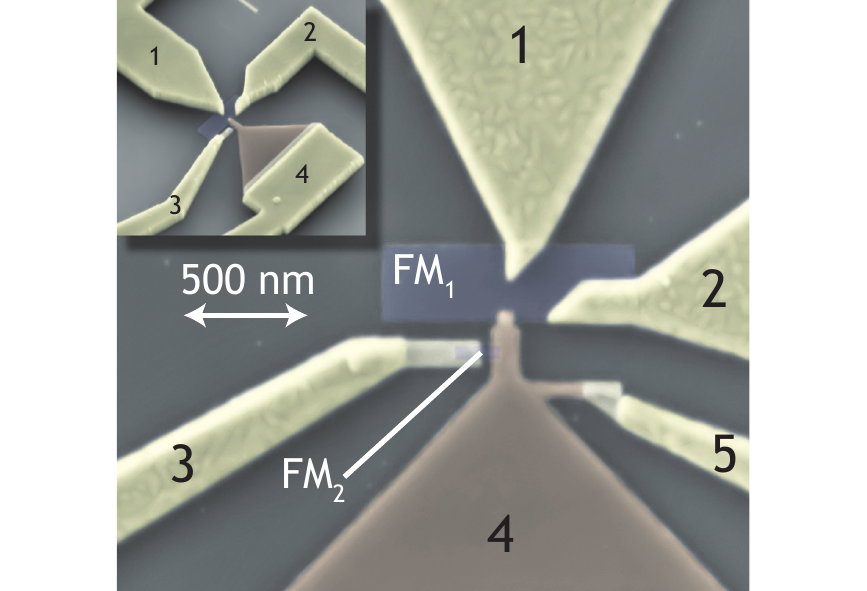}
\caption{\label{fig:3}\textbf{Coloured SEM picture of the fabricated device}. The device consists of two 15 nm thick Permalloy (Ni$_{80}$Fe$_{20}$) ferromagnets FM$_{1}$ and FM$_{2}$ of 1 $\mu$m x 300 nm and 150 x 40 nm$^2$ separated from each other by 100 nm. They are connected by a 60 nm thick copper funnel with small effective FM/NM contact areas of 40 x 40 nm$^2$ and 30 x 40 nm$^2$. 5/175 nm thick Ti/Au contacts 1 and 2 are placed asymmetrically on FM$_{1}$ to Joule heat it while contacts 3 and 4 are used to measure Joule heating and thermal spin injection. An additional contact 5 is present to measure a regular non-local spin valve signal.}
\end{figure}

To quantify the thermal injection of spins we consider the FM/NM interface and solve the Valet-Fert equation\cite{ValetFert} in a fashion similar as van Son et al\cite{vSonPRL} (Supplementary Information A). The result is depicted in figure \ref{fig:2}. A spin accumulation appears at the interface driven by the abrupt change of spin current going from the bulk FM to the bulk NM, thereby acting as an effective source of spins at the interface. The resulting spin accumulation has the following expression:

\begin{equation}\label{eq:spinacc}
\frac{\mu_{s}}{\nabla T_{FM}}= e\lambda_{F}S_{s} R_{mis}
\end{equation}

\noindent where $R_{mis}=R_{N}/(R_{N}+R_{F}/(1-P^{2}))$ is a conductivity mismatch\cite{condmism} factor in which $R_{i}=\lambda_{i}/\sigma_{i}$ are the spin resistances determined by the relaxation lengths $\lambda_{i}$ and the conductivities $\sigma_{i}$. For the metallic interfaces under consideration in this paper, this factor is very close to 1. The resulting spin accumulation induced by the heat flow $Q = - k_{FM} \nabla T_{FM}$ is determined solely by the spin Seebeck coefficient S$_{s}$ and the ferromagnetic spin relaxation length $\lambda_{F}$. Its direction is determined by the sign of the spin Seebeck coefficient which changes sign when the magnetization of the ferromagnet reverses.

\begin{figure}[b]
\includegraphics[width=8cm,keepaspectratio=true]{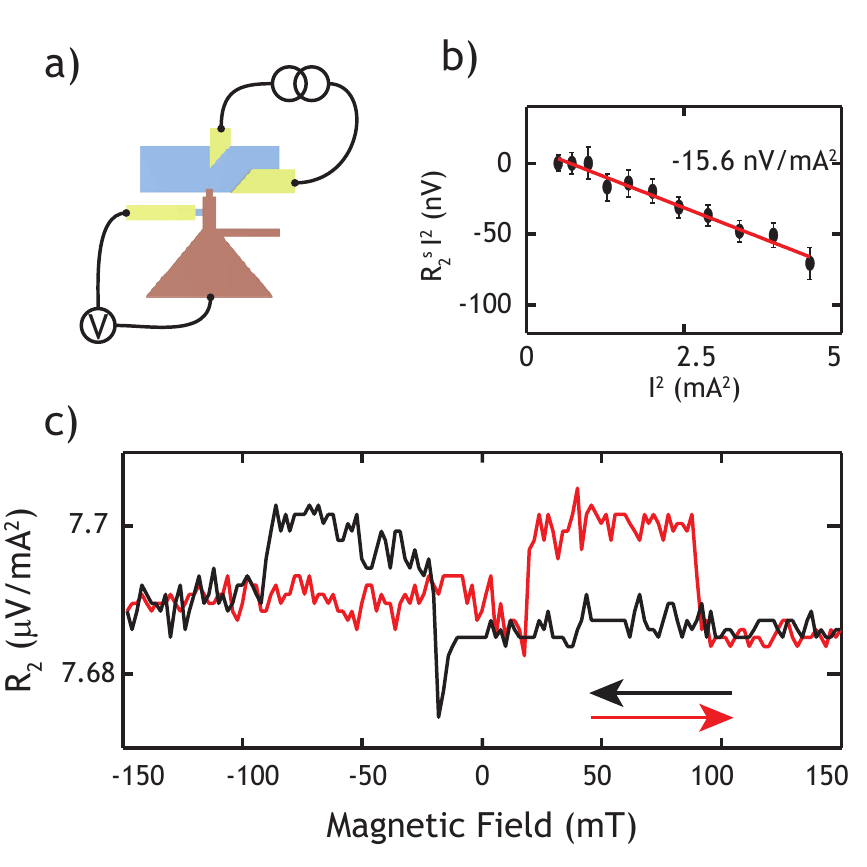}
\caption{\label{fig:4}\textbf{Thermal spin injection measurements}. \textbf{a}, Measurement scheme of the experiment. \textbf{b}, Second harmonic measurement result $R_{2}^{s} I^{2}$ (nV) of the observed thermal spin signal as a function of I$^{2}$. \textbf{c}, Measured second harmonic signal at a rms current of 1.5 mA showing the four distinct switches resulting from the magnetization alignment of FM$_{1}$ and FM$_{2}$ illustrating thermal spin injection.}
\end{figure}

Using a lock-in technique we determine the relevant parameters $R_{1}(\mu V/mA)$ and $R_{2}(\mu V/mA^{2})$ from the observed voltage\cite{FLBakker}:

\begin{equation}\label{eq:R1R2factors}
V=R_{1}I+R_{2}I^{2}+...
\end{equation}

\begin{figure}[b]
\includegraphics[width=8cm,keepaspectratio=true]{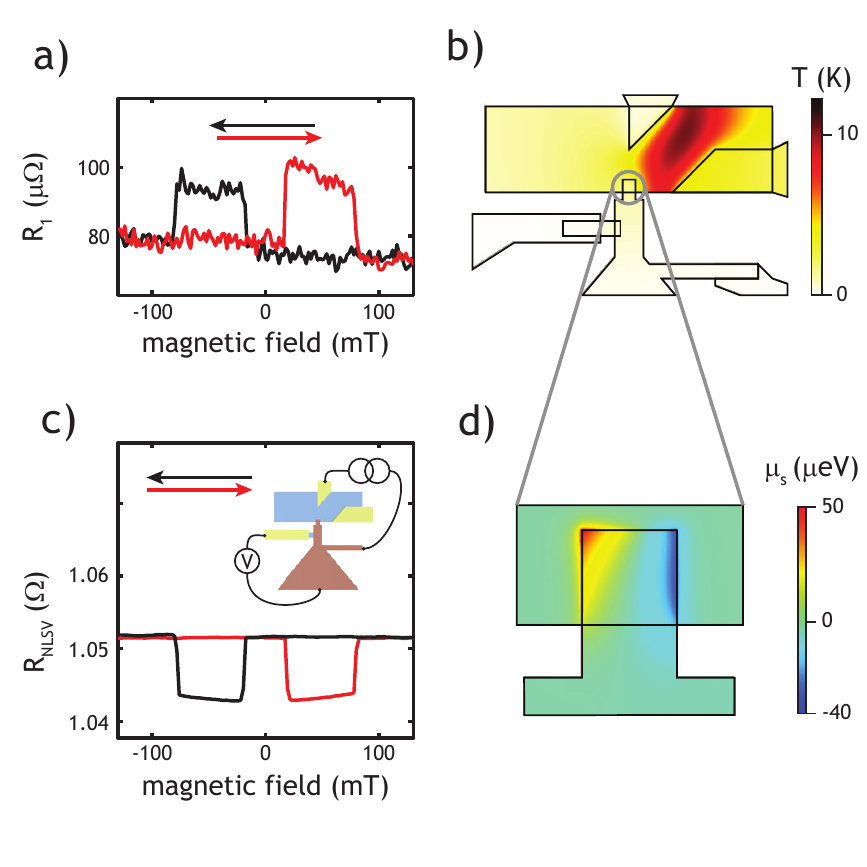}
\caption{\label{fig:5}\textbf{Rectification effects and electrical spin injection}. \textbf{a}, first harmonic measurement R$_{1}$ for the measurement setup of figure \ref{fig:4}. \textbf{b}, Calculated temperature distribution at a height of 10 nm with a current of 2 mA sent through FM$_{1}$. It illustrates the localized Joule heating, Peltier cooling and heating of the two Au/Py current injecting contacts and subsequent thermal conduction towards the three connected metallic contacts. \textbf{c}, Measured electrical spin injection scheme and resulting spin valve. \textbf{d}, Calculated spin accumulation at a height of 10 nm. A small part of the current path is short circuited by the Cu connection so that 4\% still flows in and out of the Py$_{1}$/Cu contact because of its large conductivity. This creates a large positive and negative spin accumulation. Due to the asymmetry in spin injection and the asymmetrical placement at FM$_{2}$ a small fraction of 3\% is still predicted to give a small regular spin valve signal R$_{1}^{S}$.}
\end{figure}

The baseline resistance, defined in terms of a parallel and antiparallel contribution as $(R_{i}^{P}+R_{i}^{AP})/2 = R_{i}^{b}$, allows to extract the magnitude of Joule and Peltier heating effects and possible conventional Ohmic potential drops\cite{FLBakker}. Here $R_{1}^{b}$ is determined by the Ohmic potential drop and Peltier heating/cooling measured by the FM$_{2}$-NM thermocouple while the baseline resistance $R_{2}^{b}$ is determined by Joule heating measured by the same thermocouple. The spin dependent contribution $R_{i}^{P}-R_{i}^{AP} = R_{i}^{s}$ to $R_{1}$ is due to a conventional spin valve signal while this contribution to $R_{2}$ comes from thermal spin injection.

A dedicated device was fabricated to study this effect and is shown in figure \ref{fig:3}. The heating of FM$_{1}$ has been kept very localized to an area of 150 x 150 nm$^2$ by using thick gold contacts. Moreover, the contacts are placed asymmetrically to minimize the possible current flowing in and out of the FM$_{1}$/NM interface. An additional contact 5 is present to be able to send a current directly through the FM$_{1}$/NM interface. By comparing the obtained signal $R_{1}^{s}$ to a model (see methods), we can extract the spin injection/detection efficiency\cite{FLBakker}, which has been made as high as possible by keeping the size of the FM/NM contacts small. All measurements are performed at room temperature.

Figure \ref{fig:4} shows our principal results on thermal spin injection. Four distinct P-AP and AP-P switches are observed up to 70 nV in magnitude scaling with I$^{2}$ on a large background originating from the Py$_{2}$/Cu thermocouple. 

The interpretation of the obtained signals requires a detailed knowledge of the heat, charge and spin currents in the device. For this purpose a 3D thermoelectric spin model was constructed which extends the spin-dependent current model\cite{ValetFert} to include thermoelectricity as well as thermal spin injection by the spin Seebeck coefficient. 

The calculated average contribution $R_{2}^{b}$ is 2.4 $\mu V /mA^{2}$ slightly lower then the observed 7.69 $\mu V /mA^{2}$. The difference between the observed and modelled value was seen before in non-local spin valve samples\cite{FLBakker}. It can be explained by a reduction in the Permalloy thickness due to its oxidation, which effectively increases the Joule heating. In the following, we scale the overall Joule heating in our model to fit our measured result $R_{2}^{b}$. We then find that we were able to heat FM$_{1}$ to a maximum of $\approx$ 40K at which $\nabla T_{FM}$ at the FM$_{1}$/NM interface is $\approx$ 50 K/$\mu$m. At this moment the current density is $\approx$ 8$\times$10$^{11}$ A/m$^{2}$, close to the point where the device will fail due to electromigration. 

Electrical spin injection was also measured and the results are shown in figure \ref{fig:5}c. A relatively large 9 m$\Omega$ spin valve signal is present on top of a 1.05 $\Omega$ background, being only slightly different to the 7.8 m$\Omega$ and 640 m$\Omega$ calculated signals with the metallic spin parameters $\lambda_{Cu}$ = 350 nm, $\lambda_{Py}$ = 5 nm and P$_{Py}$ = 0.25 obtained from previously fabricated samples\cite{Jedema,FLBakker}. Here P$_{Py}$ is positive as shown before\cite{Vlaminck}.

The observed thermal spin injection signal R$_{2}^{s}$ = -15.6 nV/mA$^{2}$ is determined from figure \ref{fig:4}b. We obtain a spin Seebeck coefficient for Permalloy of -3.8 $\mu V/K$, a fraction of the conventional Seebeck coefficient  $S_{F}$ = -20 $\mu V/K$\cite{Saitoh}. This gives a polarization of the Seebeck coefficient of P$_{S}$ = S$_{s}$/S$_{F}$ = 0.19 not too different from the spin polarization of the current. It is significantly larger then the S$_{s}$ = -2 $n V/K$ coefficient extracted by Uchida et al.\cite{Saitoh}. However, it is in good agreement with recent theoretical predictions \cite{HatamiPRL, Tulapurkar}.

In addition to the thermal spin injection signal, a small regular spin valve signal R$_{1}^{s}$ = -20 $\mu\Omega$ is also present and is shown in figure \ref{fig:5}a. The baseline resistance R$_{1}^{b}$ of 90 $\mu\Omega$ is in line with the calculated 95 $\mu\Omega$. This is caused by Peltier heating and cooling of the two current injecting contacts\cite{FLBakker}.

The negative regular spin valve signal R$_{1}^{s}$ can be understood as follows. Due to the high conductivity of the copper, a fraction of the current flows into and out of Cu/Py$_{1}$ interface electrically injecting spins. A small net spin accumulation at the detector interface remains caused by the asymmetric placement of FM$_{2}$. It is illustrated by the calculation of the spin accumulation at the Py$_{1}$/Cu interface shown in figure \ref{fig:5}d which shows the high geometrical dependence of this effect. The observed R$_{1}^{s}$ is somewhat smaller than the calculated -45 $\mu\Omega$. We believe that the small 40 x 40 nm$^{2}$ size of the copper contact makes sure copper grain size, lithographic precision and ballistic effects start dominating the magnitude of this effect.

A previous device showed a thermal spin injection signal -5nV/mA$^{2}$ at a FM-FM distance of 400 nm, only visible at the highest current (Supplementary information B). A similar calculation gives $S_{s}^{Py}$ = -5 $\mu V/K$ in agreement with the analysis for the sample presented.

Now that the parameters governing equation \ref{eq:spinacc} are known we may compare this to the electrical spin injection results for the transparent Cu/Py interface. We can calculate that for thermal spin injection $\mu_{S} / \nabla T \approx 2\times 10^{-14}$ eV m/K versus $\mu_{S} / J$ $\approx 3\cdot10^{-16}$ eV m$^{2}$/A for electrical spin injection through a transparent Cu/Py contact. 

Due to the lateral non-local geometry and Joule heating method used in this paper, we are limited to a maximum temperature gradient of $\approx$ 50 K/$\mu$m. However, in a typical perpendicular geometry switching by spin transfer torque\cite{CIMS} this does not have to be the case. In order to switch the magnetization by electrical spin transfer torque one needs a typical charge current density of $\approx$ 5$\cdot$10$^{11}$ A/m$^{2}$\cite{CIMS}. The same stack should be able to switch by applying a temperature difference of only a few tens of degrees as earlier theoretical studies have indicated\cite{HatamiPRL,Tulapurkar}. This simple example shows that despite the weak signals observed in this paper, this process can be a very viable alternative, or even work alongside, electrical spin injection.

In conclusion we have demonstrated the concept of thermal spin injection in a lateral metallic Py/Cu system. By constructing a theory for thermal spin injection and comparing the observed potential to detailed 3D modeling, we were able to extract a spin Seebeck coefficient for Permalloy of -3.8 $\mu V/K$. This demonstrates thermal spin injection is a feasible alternative for electrical spin injection in for example spin transfer torque experiments\cite{HatamiPRL}.

\section{Acknowledgements}
We would like to acknowledge J.G. Holstein, B. Wolfs and S. Bakker for technical assistance and T. Banarjee, P. Zomer and S. Denega for reading the manuscript. This work was financed by the European EC Contracts No. IST-033749 "DynaMax", the "Stichting voor Fundamenteel Onderzoek der Materie" (FOM) and NanoNed.

\section{Methods}

\subsection{Fabrication}
The sample in this paper was fabricated by a 1 step optical and 5 step electron beam lithography process. In each step, metals are deposited using e-beam deposition. For the e-beam lithography process a PMMA 950K resist is used of 70-400 nm thickness depending on the thickness of the deposited material and resilience to Ar ion milling. The first e-beam lithography process produces 5/30 nm thick and 100 nm wide Ti/Au markers which using an automatic alignment procedure can be aligned to in the next e-beam deposition steps with high precision. In the next four steps, the 15 nm Py, 5/30 nm Ti/Au, 5/180 nm Ti/Au and 65 nm Cu layers are deposited. For the last three steps, Ar ion milling was used prior to deposition to remove any polymer residue and the Py oxide to obtain our highly ohmic contacts.
\subsection{Measurements}
The measurements were performed using a AC current source of a frequency $<$ 1kHz far below the characteristic thermoelectric time scale of such sized systems of $\approx$ 1-100 ns. The obtained signal is sent to 3 Lock-in systems measuring the 1$^{st}$, 2$^{nd}$ and 3$^{rd}$ harmonic response simultaneously. Care was taken in deriving R$_{1}$, R$_{2}$ and R$_{3}$ by scanning the current to make sure that even higher harmonics were negligible. 
\subsection{Modeling}
We constructed a 3D model of the fabricated sample using the finite element program Comsol Multiphysics. The physics is defined in terms of a thermoelectric spin model where the spin up, down and heat currents are given by:
 
\begin{equation}\label{eq:3Dcurrentmatrix}
\left( \begin{array}{c} \vec{J_{\uparrow}} \\ \vec{J_{\downarrow}} \\ \vec{Q} \end{array} \right) =
-\left( \begin{array}{ccc} \sigma_{\uparrow} & 0 & -\sigma_{\uparrow}S_{\uparrow} \\ 0 & \sigma_{\downarrow} & -\sigma_{\downarrow}S_{\downarrow} \\ -\sigma_{\uparrow}\Pi_{\uparrow} & -\sigma_{\downarrow}\Pi_{\downarrow} & k \end{array} \right) \cdot
\left( \begin{array}{c} \vec{\nabla}\mu_{\uparrow}/e \\ \vec{\nabla}\mu_{\downarrow}/e \\ \vec{\nabla} T \end{array} \right)
\end{equation}

\noindent where $\Pi_{\uparrow,\downarrow}$ are the spin dependent Peltier coefficients given by S$_{\uparrow,\downarrow}\cdot$ T$_{0}$. Here T$_{0}$ = 300K which is the reference temperature of the device. We take these currents to be continuous across boundaries. At the end of all contacts we set the temperature to be T$_{0}$. At contact 1 in figure \ref{fig:3} we set J$_{\uparrow,\downarrow}$=J/2 to inject a charge current which is being sent through the system by setting $\mu_{\uparrow,\downarrow}$=0 at contact 2 or 5. At all other interfaces the currents are set to 0. We include Valet-Fert spin relaxation by assuming $\nabla J_{\uparrow,\downarrow}$ = $\pm\frac{(1-P^{2})\sigma_{i}}{4\lambda_{i}^{2}}(\mu_{\uparrow}-\mu_{\downarrow})$ in the bulk. Joule heating is included by assuming $\nabla Q=\zeta\ \left( \frac{\vec{J_{\uparrow}}^{2}}{\sigma_{\uparrow}}+\frac{\vec{J_{\downarrow}}^{2}}{\sigma_{\downarrow}} \right)$ where a scaling factor $\zeta=3.2$ is used to make the model correspond to the measured R$_{2}^{b}$. The system was meshed most accurately at the FM/NM interfaces where the mesh size was 1 nm in order to accurately calculate thermal spin injection. The dependencies R$_{1}^{(s)}$ up till R$_{4}^{(s)}$ were determined by calculating the results at $\pm$ 1 \& 2 mA for the parallel and antiparallel configuration. The measured resistivities $\sigma_{Au}$ = 2.2 $\cdot$ 10$^{7}$ S/m, $\sigma_{Cu}$ = 4.26 $\cdot$ 10$^{7}$ S/m and $\sigma_{Py}$ = 4.32 $\cdot$ 10$^{6}$ S/m were taken as inputs for the model. In this model, the substrate was also taken into account\cite{FLBakker}. The Seebeck coefficients $S_{Au}$=1.7 $\mu$V/K, $S_{Cu}$=1.6 $\mu$V/K, $S_{Py}$ = -20 $\mu$V/K and thermal conductances k$_{Au}$ = 300 W/m/K, k$_{Cu}$ = 300 W/m/K, k$_{Py}$ = 30 W/m/K, k$_{substrate}$ = 1 W/m/K were taken from various sources in literature\cite{Saitoh,Kittel}.

\section{Supplementary information A}

\begin{figure}[b]
\includegraphics[width=8.8cm,keepaspectratio=true]{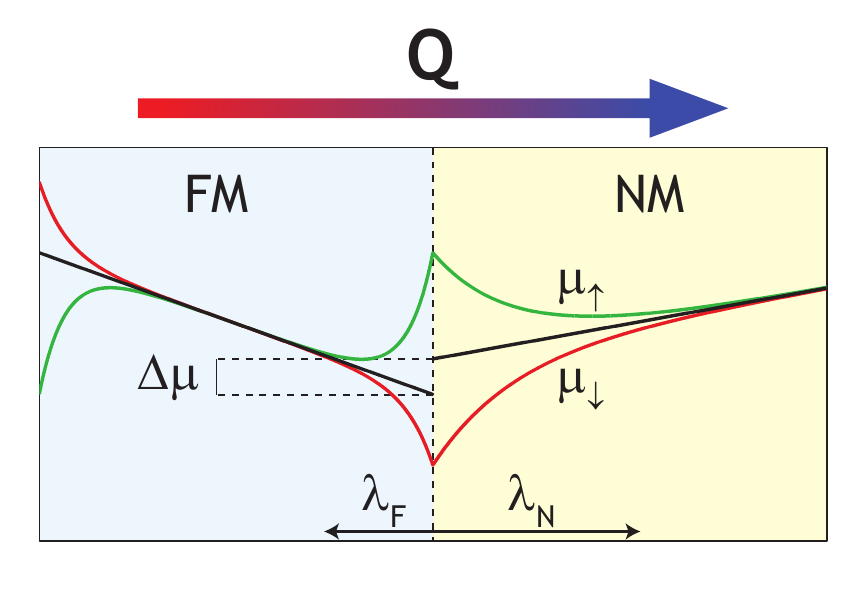}
\caption{\label{fig:1s}\textbf{Thermal spin injection by the spin Seebeck coefficient across a FM/NM interface}. Schematic figure showing the resulting spin dependent chemical potentials $\mu_{\uparrow,\downarrow}$ across a FM/NM interface when a heat current $Q=-k\nabla T$ crosses it. Heat current is taken to be continuous across the interface leading to a discontinuity in $\nabla T$. No currents are allowed to leave the FM, nevertheless, a spin current proportional to the spin Seebeck coefficient flows through the bulk FM which needs to become unpolarized in the bulk NM. This injects a spin imbalance $\mu_{\uparrow}-\mu_{\downarrow}$ at the boundary which relaxes in the FM and NM with the length scale of their respective spin relaxation lengths $\lambda_{i}$. A thermoelectric interface potential $\Delta \mu=P\mu_{s}$ also builds up\cite{vSonPRL,ValetFert}. On the left side no spin current is allowed to leave leading to an opposite spin accumulation.}
\end{figure}

Here we calculate in more detail what happens when heat is sent through the FM/NM system in figure \ref{fig:1s}. We begin by writing the spin dependent currents:

\begin{equation}\label{eq:currents2}
J_{\uparrow,\downarrow}=-\sigma_{\uparrow,\downarrow}(\frac{1}{e}\nabla \mu_{\uparrow,\downarrow}-S_{\uparrow,\downarrow}\nabla T)
\end{equation}

\noindent here $\mu_{\uparrow,\downarrow}$ is the spin dependent chemical potential. When a heat current Q is send through a bulk ferromagnet in the absence of a charge current, a spin current $J_{s}=J_{\uparrow}-J_{\downarrow}=\sigma_{F}(1-P^{2})S_{s}\nabla T/2$ flows, driven by the spin Seebeck coefficient $S_{s}=S_{\uparrow}-S_{\downarrow}$. Here $P$ is the current polarization of the FM and $\sigma_{F}$ is the conductivity of the ferromagnet. Charge and spin current conservation\cite{vSonPRL,ValetFert} leads to the thermoelectric spin diffusion equation:

\begin{equation}\label{eq:diffeq2}
\nabla^2 \mu_{s}=\frac{\mu_{s}}{\lambda^{2}}+e(\frac{dS_{s}}{dT}(\nabla T)^{2}+S_{s}\nabla^2 T)
\end{equation}

\noindent where $\mu_{s}$ is the spin accumulation $\mu_{\uparrow}-\mu_{\downarrow}$. In addition to the Valet-Fert spin diffusion equation $\nabla^2 \mu_{s}=\frac{\mu_{s}}{\lambda^{2}}$ two source terms are present. Both terms can create (albeit small) bulk spin accumulations. 

\begin{figure}[b]
\includegraphics[width=8.8cm,keepaspectratio=true]{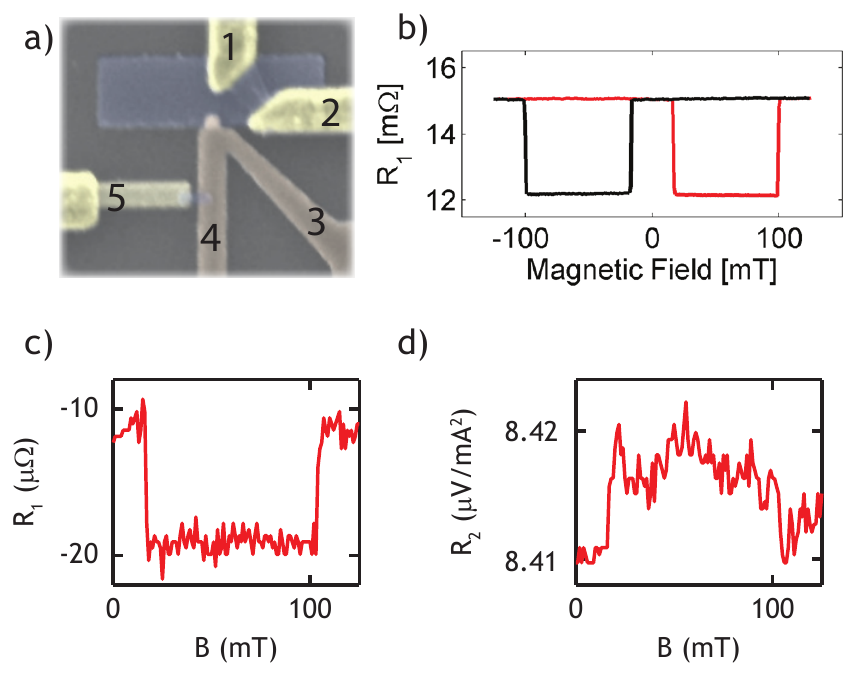}
\caption{\label{fig:2s}\textbf{Previous device results}. a) Coloured SEM figure of the device. The sample consist of the same two ferromagnets which are now placed 400 nm apart. It is connected by a copper V shape instead of a funnel. b) Non local spin valve signal by sending current from contact 1 to 3 and measuring the potential from contact 5 to 4. c,d) Thermal spin injection result. The current is now send from contact 1 to 2 while the potential was measured between contacts 5 and 4.}
\end{figure}

In figure \ref{fig:1s} we send a heat current Q through the FM/NM interface while we allow no charge or spin current to leave. The heat current $Q=-k\nabla T$ needs to be continuous throughout the system, leading to $\nabla T_{FM} = k_{NM}/k_{FM} \nabla T_{NM}$ at the interface. Since $\nabla T$ is constant in both regions individually, and for first order effects we may assume $S_{s}$ is constant, the source terms in equation \ref{eq:diffeq2} are irrelevant. Therefore, we may use the standard Valet-Fert spin diffusion equation to solve the bulk spin accumulation leading to the general expression for the spin dependent potentials in the bulk: 

\begin{equation}\label{eq:diffeq3}
\mu_{\uparrow,\downarrow}(x) = A + Bx \pm C/\sigma_{\uparrow,\downarrow}e^{-x/\lambda_{i}} \pm D/\sigma_{\uparrow,\downarrow} e^{x/\lambda_{i}}
\end{equation}

\noindent with A-D the parameters to be solved in both regions. At the FM/NM interface we take the chemical potentials $\mu_{\uparrow,\downarrow}$ to be continuous as well as the spin dependent currents $J_{\uparrow,\downarrow}$. At the outer interfaces we set the spin dependent currents to zero. This leads to a set of equation which can be solved. We obtain:

\begin{equation}\label{eq:SFM}
B=e\frac{\sigma_{\uparrow}S_{\uparrow}+\sigma_{\downarrow}S_{\downarrow}}{\sigma_{\uparrow}+\sigma_{\downarrow}} \nabla T_{FM}\equiv e S_{FM} \nabla T_{FM}
\end{equation}

\noindent where we use the definition of the conventional Seebeck coefficient of a ferromagnet $S_{FM}$\cite{TEP72}. The spin accumulation at the interface is:

\begin{equation}\label{eq:spinaccumulation}
\frac{\mu_{s}}{\nabla T_{FM}}= e\lambda_{F}S_{s} R_{mis}
\end{equation}

\noindent where $R_{mis}=R_{N}/(R_{N}+R_{F}/(1-P^{2}))$ is a conductivity mismatch\cite{condmism} factor in which $R_{i}=\lambda_{i}/\sigma_{i}$ are the spin resistances determined by the relaxation lengths $\lambda_{i}$ and the conductivities $\sigma_{i}$. 

\section{Supplementary information B}

A SEM picture of the previous sample is shown in figure \ref{fig:2s} (a). A regular spin valve signal was measured by sending a current from contact 1 to 3 and measuring the potential between contact 5 and 4 of which the result is shown in figure \ref{fig:2s} (b). In this case a 13.8 m$\Omega$ background R$_{1}^{b}$ is observed on top of a non local spin valve signal R$_{1}^{s}$ of 3 m$\Omega$. The background is originating from Peltier heating/cooling of the FM/NM interfaces\cite{FLBakker}. Both signals are close to the calculated 14.1 m$\Omega$ and 4.1 m$\Omega$. 

When the current is send from contact 1 to 2, we obtain the results shown in figure \ref{fig:2s} (c,d).  A regular spin valve signal R$_{1}^{s}$ of 10 $\mu\Omega$ is observed on top of a small -15 $\mu\Omega$ background R$_{1}^{b}$. This is somewhat different then the calculated -100 $\mu\Omega$ background and -4 $\mu\Omega$. However, these effects are highly dependent on the exact geometry and are due to the small 30 x 30 nm$^{2}$ size of the contact. This makes sure grain size, lithographic precision and ballistic effects dominate. 

Thermal spin injection was observed and is shown in figure \ref{fig:2s} (d). The background R$_{2}^{b}$ is again larger then the calculated 3.4 $\mu V/mA^{2}$. If we compensate for this in the modelling we obtain from the observed $\approx$ -7 nV/mA$^{2}$ signal a spin Seebeck coefficient for Permalloy of $\approx$ -5 $\mu V/K$.

We note that our thermal spin injection signals $R_{2}^{s}I^{2}$ cannot be due to an induced temperature dependence of $R_{1}^{s}I$ because $R_{3}^{s}I^{3}$, representing the Joule heating induced change of $R_{1}^{s}I$ should be $\approx$10 times larger then the possible Peltier heating induced $R_{2}^{s}I^{2}$ at the currents we used in both samples. However, $R_{3}I^{3}$ was simultaneously measured and found absent. Furthermore, The signal $R_{1}^{s}I$ of both devices is of different sign, excluding any further dependence of $R_{2}^{s}I^{2}$ on $R_{1}^{s}I$.

We conclude that also in this device we have good agreement between observed and calculated thermoelectric voltages when we apply a similar correction for the Joule heating. A very similar value for the spin Seebeck coefficient was found.

\end{document}